\documentclass[11pt]{article}
\usepackage{amsmath,amsthm,amssymb,amscd}
\usepackage{hyperref}
\usepackage{geometry}
\usepackage{tikz-cd}
\usepackage{booktabs} % For better table aesthetics
\usepackage{graphicx} % For including images
\usepackage{multirow}
\usepackage{tikz}
\usetikzlibrary{calc,positioning,3d}
\usepackage[normalem]{ulem} % cross through
\usepackage{imakeidx}
\usepackage{makecell}

\makeindex

\usepackage{subfigure} % For subfigures

\theoremstyle{definition}

\title{CAML: Commutative algebra machine learning --- a case study on protein-ligand binding affinity prediction}
		
	\author{Hongsong Feng$^1$,  
	Faisal Suwayyid $^{~2,3}$, 
 	Mushal Zia$^3$,
	JunJie Wee$^3$, 
	Yuta Hozumi\footnote{Current address: School of Mathematics, Georgia Institute of Technology, Atlanta, GA, USA.} ~$^3$\\
   Chunlong Chen$^4$ 
		and Guo-Wei Wei\footnote{Corresponding author: Guo-Wei Wei (weig@msu.edu).~}$^{~3,5,6}$ \\		
		$^1$Department of Mathematics and Statistics,\\
		University of North Carolina at Charlotte, Charlotte, NC 28223, USA\\
		$^2$Department of Mathematics,\\
		King Fahd University of Petroleum and Minerals, Dhahran 31261, KSA.\\
		$^3$Department of Mathematics,\\
		Michigan State University, MI 48824, USA.\\
		$^4$Physical Sciences Division, \\Pacific Northwest National Laboratory, Richland, Washington 99354, USA. \\
		$^5$Department of Electrical and Computer Engineering,\\
		Michigan State University, MI 48824, USA.\\
		$^6$Department of Biochemistry and Molecular Biology,\\
		Michigan State University, MI 48824, USA.
	}

	\date{\today}
\begin{document}
	\maketitle 
\begin{abstract}	    
Recently, Suwayyid and Wei have introduced commutative algebra as an emerging paradigm for machine learning and data science. In this work, we integrate commutative algebra machine learning (CAML) for the prediction of protein-ligand binding affinities. Specifically, we apply persistent Stanley–Reisner theory, a key concept in combinatorial commutative algebra, to the affinity predictions of protein-ligand binding and metalloprotein-ligand binding. We introduce three new algorithms, i.e., element-specific commutative algebra, category-specific commutative algebra, and commutative algebra on bipartite complexes, to address the complexity of data involved in (metallo) protein-ligand complexes. We show that the proposed CAML outperforms other state-of-the-art methods in (metallo) protein-ligand binding affinity predictions.   
 \end{abstract}
	
Keywords: Persistent commutative algebra, facet persistence barcodes,  persistent ideals, machine learning, protein-ligand binding.     
	
{\setcounter{tocdepth}{4} \tableofcontents}
	 \setcounter{page}{1}
	 \newpage	
	
\section{Introduction}
Drug discovery plays a vital role in contemporary medicine, profoundly impacting global health outcomes. Conventional drug development processes, however, are time-intensive and costly, requiring more than a decade and billions of dollars to commercialize a single drug \cite{fleming2018computer}. Established techniques such as molecular docking \cite{lyu2019ultra,kitchen2004docking,pinzi2019molecular,pagadala2017software}, free energy perturbation \cite{wang2015accurate}, and empirical modeling \cite{sliwoski2014computational} have propelled advancements but face inherent constraints. These methods often suffer from inaccuracies, demand substantial computational resources for large-scale analyses, and may overlook novel binding sites or interaction dynamics, potentially missing therapeutic breakthroughs.

Machine learning (ML) approaches are gaining traction as powerful tools in drug design \cite{jumper2021highly,baek2021accurate,lin2023evolutionary,song2024multiobjective}, renowned for their capacity to forecast protein structures and detect intricate patterns for enhanced predictions \cite{luo2019challenges}. The adoption of deep learning, integrated with chemoinformatics and bioinformatics \cite{lo2018machine}, marks a transformative shift toward data-driven methodologies in pharmaceutical research \cite{atomwise2024ai,gomez2024inactive,hu2022discovery}. Nevertheless, challenges persist, including limited datasets \cite{dou2023machine}, data imbalance \cite{jiang2025review}, intricate molecular architectures, and stereochemical complexities. Furthermore, embedding essential physical interactions, such as hydrogen bonding, van der Waals forces, hydrophobic effects, electrostatic forces, and ionic bonds, into ML algorithms for protein-ligand binding remains a significant hurdle \cite{ballester2010machine,wang2017improving,li2023development}.

To address these limitations, researchers are employing sophisticated mathematical frameworks rooted in algebraic topology, differential geometry, and combinatorial graph theory \cite{nguyen2020review}. These multiscale models, previously successful in characterizing biomolecular systems \cite{cang2018representability,nguyen2019mathematical,wang2020persistent,meng2021persistent,chen2024multiscale}, capture fundamental physical, chemical, and biological interactions critical to protein-ligand binding while clarifying the 3D structural intricacies of these complexes. Notably, these approaches have delivered top-tier performances in the D3R Grand Challenges, a leading international competition in computer-aided drug design \cite{nguyen2019mathematical, nguyen2020mathdl}. Inspired by this success, there is a continuous effort in computational biology and applied mathematics to seek advanced mathematical representations of complex biomolecules, such as proteins and their interactions.
	  
Commutative algebra is a branch of mathematics that studies commutative rings, their ideals, modules, and related structures \cite{miller2005combinatorial,eisenbud2013commutative}. It serves as a foundational framework for algebraic geometry, number theory, and many other areas in mathematics. Its key concepts include Noetherian rings, Cohen-Macaulay rings, localization theory, primary decomposition, dimension theory, and homological algebra.

Despite its importance in pure mathematics, it has hardly been applied to data science and artificial intelligence. Recently, Suwayyid and Wei introduced persistent Stanley-Reisner theory to bridge commutative algebra, algebraic topology, machine learning, and data science \cite{suwayyid2025persistent}. Stanley-Reisner theory is the study of the commutative algebra, i.e., square-free monomial ideals in a polynomial ring, of simplicial complexes, structured sets comprising points, line segments, triangles, and their higher-dimensional counterparts \cite{stanley1996combinatorics, bruns1998cohen,ha2008monomial}. Therefore, persistent Stanley-Reisner theory (PSRT) enables commutative algebra analysis (CAA) of point cloud data and machine learning predictions. Specifically, PSRT examines how the Stanley-Reisner structure of a simplicial complex evolves under filtration. Many computable quantities, including persistent graded Betti numbers via Hochster’s formula, persistent 
$f$-vectors, persistent $h$-vectors, and persistent facet ideals, have been proposed. Facet persistence barcodes, which record the birth and death of persistent facet ideals as the simplicial complex evolves, have been introduced for practical applications in data science.
PSRT provides novel insights into geometry, topology, and combinatorics at multiple scales.
An important motivation for this development was persistent homology \cite{edelsbrunner2008persistent, zomorodian2005computing}, an algebraic topology tool for topological data analysis (TDA), and topological deep learning (TDL) \cite{cang2017topologynet}, a new frontier for relational learning \cite{papamarkou2024position}.

The objective of this work is to explore the utility and demonstrate the potential of commutative algebraic machine learning (CAML) for protein-ligand binding affinity prediction. We consider two benchmark datasets: the PDBbind-v2016 dataset for protein-ligand binding interactions \cite{liu2015pdb} and a metalloprotein-ligand binding dataset \cite{jiang2023metalprognet}. As these datasets involve intricate three-dimensional (3D) protein-ligand complexes as well as complex physical and chemical interactions, we propose a few new CAML algorithms, i.e., element-specific commutative algebra, category-specific commutative algebra, and commutative algebra on bipartite complexes, to capture intrinsic physical and chemical interactions, such as hydrogen bonding, van der Waals forces, hydrophobic effects, electrostatic forces, and ionic bonds in 3D metalloprotein-ligand complexes. As shown in the results, the proposed CAML consistently outshines its peers, achieving state-of-the-art outcomes across benchmark datasets in protein-ligand binding affinity prediction.

The rest of this paper is organized as follows. Section~2 is devoted to the results of CAML for protein-ligand binding and metalloprotein-ligand binding predictions. Methods are described in Section~3. This paper ends with a conclusion.
 
\section{Results} 
\subsection{Protein-ligand binding affinity predictions}

The PDBbind database \cite{liu2015pdb} is a widely recognized, curated resource that systematically collects experimentally determined 3D structures of protein-ligand complexes alongside their corresponding binding affinity data e.g., dissociation constants 
$K_d$, inhibition constants $K_i$, and Gibbs free energy changes $\Delta G$, It serves as a gold-standard benchmark for developing and validating computational models aimed at predicting protein-ligand binding affinities.

\begin{figure}[!htb]
	\centering
	\includegraphics[width=1\textwidth]{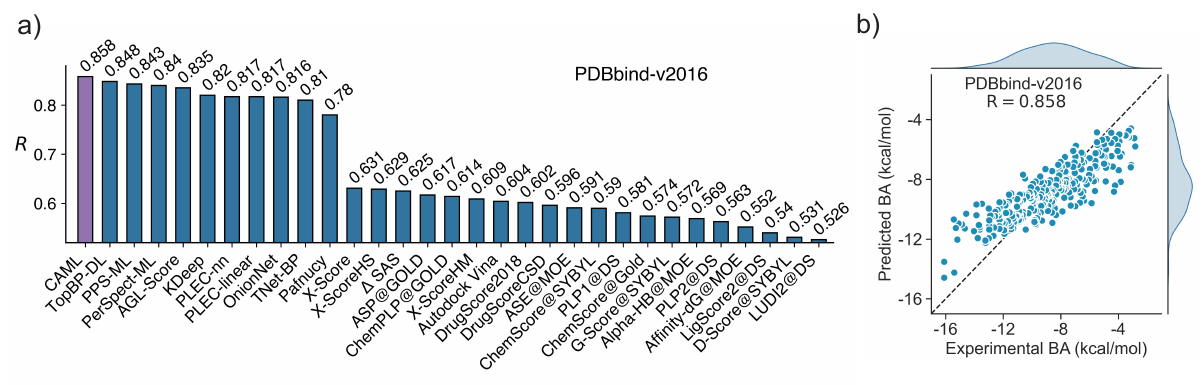}
	\caption{a. A comparison of the predictions from our CAML model with other published models in terms of Pearson correlation coefficient $(R)$ on the PDBbind-v2016 dataset. b. A comparison between the experimental and predicted binding affinities (BAs) from our CAML model for the PDBbind-v2016 dataset.}
	\label{Fig:R-comparison-v2016}
\end{figure}

Leveraging our Persistent Stanley-Rensiner Theory (PSRT), we developed commutative algebra machine learning (CAML) models to predict protein-ligand binding affinities. The models were benchmarked against established methods using the widely recognized PDBbind dataset. Specifically, we focused on the PDBbind-v2016 subset,  a rigorously curated version with clearly defined training (3,768 complexes) and test sets (290 complexes). The PDBbind database \cite{liu2015pdb} provides a comprehensive collection of 3D protein-ligand structures paired with binding affinity data. As shown in \autoref{Fig:R-comparison-v2016}a, our PSRT-guided model, CAML, outperformed existing state-of-the-art approaches, achieving superior predictive accuracy in binding affinity estimation.

Numerous competitive models rooted in mathematical or physical frameworks \cite{ballester2010machine,wang2017improving,li2023development}, such as persistent homology \cite{cang2018representability}, persistent spectral theories \cite{meng2021persistent,liu2023persistent}, have been reported. These rank among the top performers in this domain (see \autoref{Fig:R-comparison-v2016}a). On the PDBbind-v2016 test set, CAML achieved a Pearson correlation coefficient ($R$) of 0.858, significantly surpassing persistent homology-based TopBP-DL ($R$=0.848)\cite{cang2018representability} and persistent spectral theory-based models PerSpect-ML ($R$=0.843)\cite{meng2021persistent} and PPS-ML ($R$=0.840)\cite{liu2023persistent}. These results highlight CAML’s efficacy as a novel analytical tool and its ability to drive advanced predictive models for binding affinity.

CAML’s reliability is further demonstrated by the strong alignment between experimental and predicted binding affinities, as visualized in \autoref{Fig:R-comparison-v2016}b. Its success stems from three key innovations: (1) PSRT-driven molecular data analysis, (2) element-specific (ES) and category-specific (CS) modeling of intra- and intermolecular interactions, and (3) integration of natural language processing (NLP) via a transformer architecture (details in \autoref{sec:Vectorization}).

As detailed in \autoref{table:Pearson-values-PDBbind}, we evaluated five distinct models. The top performer (final CAML model, $R$=0.858) combines consensus predictions from CAML(ES, CS)—a fusion of element- and category-specific strategies—with transformer-based sequence analysis. The ES approach, a widely adopted method for dissecting atomic interactions, and the CS strategy, which categorizes interactions by atomic properties, were individually effective. Their combination (CAML(ES, CS), $R$=0.853) further enhanced accuracy. Integrating these with sequence-based NLP predictions via the transformer model yielded the final CAML’s performance, highlighting the synergy between structural and sequential pattern analysis.

This multi-faceted approach positions CAML as a state-of-the-art tool for protein-ligand binding affinity prediction, with implications for drug discovery and molecular design.

\begin{table}[htb!]
	\small
	\centering
	\begin{tabular}{c | c | c | c | c |c}
		\hline
		\textbf{Dataset} &  \textbf{CAML(ES)} &  \textbf{CAML(CS)} & \textbf{CAML(ES,CS)} & \textbf{Transformer} & \makecell{\textbf{CAML(ES,CS)} \\ \textbf{+ Transformer}} \\
		\hline
		PDBbind-v2016 & 0.836(1.743)& 0.834(1.745) & 0.845(1.719) & 0.836(1.713) & \textbf{0.858 (1.669)} \\
		\hline
	\end{tabular}
	\caption{Modeling performance of various strategies on the test set of PDBbind-v2016. The evaluation metrics used are the Pearson correlation coefficient ($R$) and root mean square error (RMSE, in kcal/mol). Twenty independent runs with different random seeds were performed, and the average metric values are reported. CAML(ES) and CAML(CS) refer to commutative algebraic machine learning models combined with element-specific and category-specific atom combinations, respectively. CAML(ES,CS) represents the consensus results from the CAML(ES) and CAML(CS) models. Transformer refers to sequence-based modeling using natural language processing. CAML(ES,CS)+Transformer indicates the consensus predictions from the CAML(ES,CS) and Transformer models, which is defined as our final CAML model.}
	\label{table:Pearson-values-PDBbind}
\end{table}

\subsection{Metalloprotein-ligand binding affinity predictions} 

As another benchmark example, we consider  metalloprotein-ligand binding affinities. 
Metalloproteins are proteins that incorporate metal ions as integral structural components and play indispensable roles in biological processes such as cellular respiration, electron transfer, catalytic reactions, and structural stabilization \cite{andreini2009metalloproteomes, banci2012metallomics,chalkley2022novo}. More specifically, protein metal-binding sites are responsible for catalyzing some of the most difficult and yet important functions, such as photosynthesis, respiration, water oxidation, molecular oxygen reduction, and nitrogen fixation. Studies estimate that roughly half of all proteins in biology metalloproteins \cite{valasatava2018extent,hay1984bio,waldron2009bacterial}. The  prediction of metalloprotein-ligand binding affinities represents a critical challenge in drug discovery.  Deciphering the structure, such as function relationships and interaction mechanisms of metalloproteins, is pivotal for unraveling fundamental biological pathways and accelerating the design of targeted therapeutics.

Recent advancements have addressed the scarcity of specialized datasets for this task. For example, the study by \cite{jiang2023metalprognet} introduced the largest curated dataset to date for metalloprotein-ligand binding affinity prediction, providing a robust foundation for developing and benchmarking computational models in this domain.

\begin{figure}[!htb]
	\centering
	\includegraphics[width=0.7\textwidth]{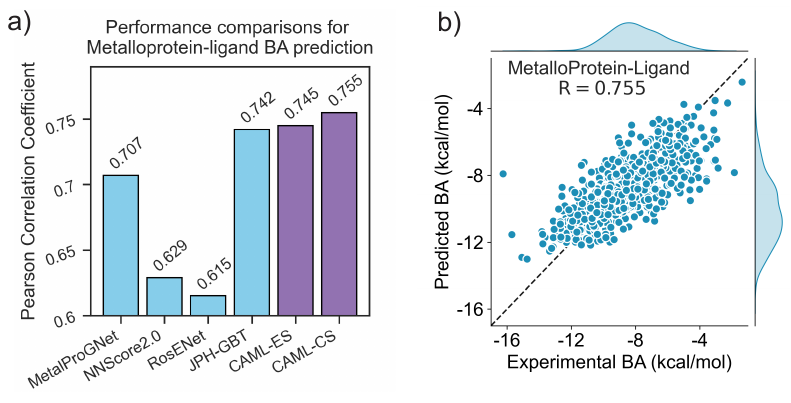}
	\caption{a. The prediction performance of our CAML models on the metalloprotein-ligand binding affinity (BA) dataset. b. A comparison between the experimental binding affinities and the predicted values from our CAML(CS) model for this dataset.}
	\label{Fig:R-comparison-metal}
\end{figure}

Using the dataset from \cite{jiang2023metalprognet}, we constructed two CAML machine learning models based on element-specific (ES) and category-specific (CS) strategies, resulting in CAML(ES) and CAML(CS) models. \autoref{Fig:R-comparison-metal}a gives the comparisons between our CAML models with other published models in terms of Pearson correlation coefficient (R). The previous state-of-art model is JPH-GBT model \cite{wang2025join}, which gave a much higher R value than other models \cite{hassan2020rosenet,durrant2011nnscore,jiang2023metalprognet}. Our CAML models redefine the state-of-art for metalloprotein-ligand binding affinity predictions. Model CAML(ES) and CAML(CS) give $R$ values of o.745 and 0.755, respectively. \autoref{Fig:R-comparison-metal}b shows that the comparison between the experimental binding affinity values and the predicted one using our CAML(CS) model. \autoref{table:metal-performance-comparison} gives the comparisons of our models with others in terms of $R$ and RMSE metrics.

\begin{table}[htb!]
	\small
	\centering
	\begin{tabular}{c |c |c }
		\hline
		\textbf{Machine learning models} &  
		\makecell{\textbf{Pearson correlation} \\ \textbf{coefficient $(R)$}} &  
		\makecell{\textbf{Root-mean-squared error } \\ \textbf{(RMSE)}} \\
		\hline
		RosENet \cite{jiang2023metalprognet,hassan2020rosenet} & 0.615 $\pm$ 0.017 & 1.436 $\pm$ 0.011\\
		NNScore2.0 \cite{jiang2023metalprognet,durrant2011nnscore}& 0.629 $\pm$ 0.002 & 1.391 $\pm$ 0.004\\
		MetalProGNet \cite{jiang2023metalprognet} & 0.703 $\pm$ 0.010 & 1.285 $\pm$ 0.020\\
		JPH-GBT \cite{wang2025join} & 0.742 $\pm$ 0.001 & 1.205 $\pm$ 0.001\\
		CAML(ES) & 0.745 $\pm$ 0.001 & 1.202 $\pm$ 0.002\\
		CAML(CS) & \textbf{0.755 $\pm$ 0.001} & \textbf{1.185 $\pm$ 0.002}\\
		\hline
	\end{tabular}
	\caption{Comparison of our CAML models with existing machine learning approaches in modeling metalloprotein-ligand binding affinity dataset. CAML(ES) and CAML(CS) refer to models utilizing the PSRT vectorization framework combined with element-specific (ES) and category-specific (CS) atom groupings, respectively. The RMSE values are computed based on the raw \( pK_d \) labels that serve as binding affinities.}
	\label{table:metal-performance-comparison}
\end{table}

\section{Methods}

In this section, we first list our datasets, showing the clear separation between training and test sets.  Then, we we provide an overview of persistent Stanley--Reisner theory (PSRT). Next, we describe the vectorization of persistent commutative algebra, following by natural language processing (NLP) molecular descriptors. Machine learning models and model parameters are given.  We also define the evaluation metrics.

\subsection{Dataset}

The first benchmark data set we use is PDBbind-v2016, which is the largest protein-ligand binding affinity prediction dataset with well-defined training and test sets. The second one is metalloprotein-ligand dataset complied in work \cite{jiang2023metalprognet}. This original dataset consists of training, validation, and test sets with different types of metal ions. We utilize their training and test to benchmark the performance of PSRT-based machine learning models. 

\begin{table}[htb!]
	\small
	\centering
	\begin{tabular}{c| c c  c }
		\hline
		\textbf{Dataset} &  \textbf{Total} & \textbf{Training set } & \textbf{Test set}\\
		\hline
		 PDBbind-v2016 \cite{liu2015pdb} & 1300& 1105 &195\\
		Metalloprotein-ligand \cite{wang2025join} & 2463  &1845& 618 \\
		\hline
	\end{tabular}
	\caption{Details of the datasets utilized for benchmark tests in this study.}
	\label{table:datasets-information}
\end{table}

\subsection{Persistent Stanely--Reisner Theory}

Persistent Stanley--Reisner theory is a novel framework for analyzing the shape of data by leveraging tools from combinatorial commutative algebra \cite{suwayyid2025persistent}. It encodes point cloud data as simplicial complexes—combinatorial structures built from vertices, edges, triangles, and higher-dimensional simplices—capturing both topological and combinatorial features inherent in the data. A filtration process is then applied to these complexes to track the evolution and persistence of such features across multiple spatial or geometric scales. This approach introduces algebraic invariants such as persistent $h$-vectors, $f$-vectors, graded Betti numbers, and facet ideals, thus providing a new algebraic perspective within the broader framework of topological data analysis.

\subsubsection{Simplicial Complex}

A \emph{simplicial complex} \(\Delta\) on the finite vertex set
\(	V = \{x_1, x_2, \dots, x_n\}\) is a collection of subsets of \(V\), referred to as \emph{faces} or \emph{simplices}, satisfying the following conditions:
\begin{enumerate}
	\item %\textbf{Hereditary property:} 
	If \(F \in \Delta\) and \(G \subseteq F\), then \(G \in \Delta\).
	\item %\textbf{Inclusion of vertices:} 
	For each \(i = 1, \dots, n\), the singleton \(\{x_i\}\) belongs to \(\Delta\). In particular, every vertex is included as a face.
\end{enumerate}
A face consisting of \(r+1\) vertices is called an \emph{\(r\)-dimensional face}. The \emph{dimension} of \(\Delta\) is defined as the maximum dimension among its faces. A face that is maximal with respect to inclusion is called a \emph{facet} of \(\Delta\), and the set of all such facets is denoted by \(\mathcal{F}(\Delta)\).

\subsubsection{Stanley--Reisner Theory and Facet ideals}

Let \( k \) be a field, and consider the standard polynomial ring
\begin{equation}\label{eq:S-def}
	S = k[x_1, x_2, \dots, x_n],
\end{equation}
endowed with the natural \(\mathbb{Z}\)-grading determined by \(\deg(x_i) = 1\) for all \(i = 1, \dots, n\). Let \(\Delta\) be a simplicial complex on the vertex set \(V = \{x_1, x_2, \dots, x_n\}\). The \emph{Stanley--Reisner ideal} associated with \(\Delta\) is defined as
\begin{equation}\label{eq:SR-ideal}
	I(\Delta) = \left\langle x_{i_1} x_{i_2} \cdots x_{i_r} \;\middle|\; \{x_{i_1}, x_{i_2}, \dots, x_{i_r}\} \notin \Delta \right\rangle,
\end{equation}
that is, it is the ideal generated by all squarefree monomials corresponding to non-faces of \(\Delta\). The quotient ring
\begin{equation}\label{eq:SR-ring}
	k[\Delta] = S / I(\Delta)
\end{equation}
is called the \emph{Stanley--Reisner ring} of \(\Delta\). 

It is a classical result that the Krull dimension of the Stanley--Reisner ring is given by
\begin{equation}\label{eq:Krull-dim}
	\dim\bigl(k[\Delta]\bigr) = \dim(\Delta) + 1,
\end{equation}
which we denote by \(d\). Thus, the simplicial complex \(\Delta\) is said to be \((d-1)\)-dimensional.

Now, for any subset \(A \subseteq V = \{x_1, x_2, \dots, x_n\}\), define the associated \emph{prime monomial ideal} by
\[
P_A := \left( x_i \mid x_i \notin A \right).
\]
In the context of Stanley--Reisner theory, we are particularly interested in the prime monomial ideals associated with the facets of \(\Delta\), referred to as the \emph{facet prime monomial ideals}, or simply, \emph{facet ideals}.

A fundamental property of the Stanley--Reisner ideal is that it admits a primary decomposition as the intersection of the facet ideals:
\[
I(\Delta) = \bigcap_{\sigma \in \mathcal{F}(\Delta)} P_\sigma,
\]
where \(\mathcal{F}(\Delta)\) denotes the set of all facets of the simplicial complex \(\Delta\).

\subsubsection{Graded Betti numbers, f-vectors and h-vectors}

As a graded \(S\)-module, \(k[\Delta]\) admits a minimal free resolution of 
the form
\begin{equation}\label{eq:min-free-res}
	\cdots 
	\;\longrightarrow\;
	\bigoplus_{j} S(-j)^{\beta_{i,j}\!\bigl(k[\Delta]\bigr)}
	\;\longrightarrow\;
	\cdots 
	\;\longrightarrow\;
	\bigoplus_{j} S(-j)^{\beta_{0,j}\!\bigl(k[\Delta]\bigr)}
	\;\longrightarrow\;
	k[\Delta]
	\;\longrightarrow\;
	0,
\end{equation}
where $ S(-j) $ is the graded free module $S$ shifted in degree by $j$ and  {\it graded Betti numbers} are
\begin{equation}\label{eq:Betti-numbers-def}
	\beta_{i,j}\!\bigl(k[\Delta]\bigr)
	\;=\;
	\dim_k \operatorname{Tor}^S_i\!\bigl(k[\Delta], k\bigr)_j,
\end{equation}
with ${\rm Tor}^S_i\!\bigl(k[\Delta], k\bigr)_j$ being the Tor module, which measures how nontrivial the resolution is at homological degree $i$.    
For a subset \(W \subseteq V\), the \emph{restriction} (or induced subcomplex) 
of \(\Delta\) to \(W\) is 
\begin{equation}\label{eq:induced-subcomplex}
	\Delta_W
	\;=\;
	\{\,
	\tau \in \Delta 
	: 
	\tau \subseteq W
	\}.
\end{equation}

Hochster's formula provides an explicit description of the \(\mathbb{Z}\)-graded Betti numbers \(\beta_{i, j+i}(k[\Delta])\) of the Stanley--Reisner ring \(k[\Delta]\) in terms of the reduced simplicial homology of induced subcomplexes. For integers \(i, j \geq 0\), it states that
\begin{equation}\label{eq:Hochster-alt}
	\beta_{i,j+i}\!\bigl(k[\Delta]\bigr)
	=
	\sum_{\substack{W \subseteq \{x_1, \dots, x_n\} \\ |W| = j+i}}
	\dim_k \widetilde{H}_{j-1}\!\bigl(\Delta_W; k\bigr),
\end{equation}
where \(\Delta_W\) denotes the subcomplex of \(\Delta\) induced on the vertex set \(W\), and \(\widetilde{H}_{j-1}(\Delta_W; k)\) is the \((j-1)\)-st reduced homology group with coefficients in \(k\). This formula holds for \(1 \leq i \leq n-1\) and \(1 \leq j \leq \min\{n - i, \dim(\Delta) + 1\}\).

In particular, for \(j = 1\), the formula simplifies to
\begin{equation}\label{eq:Hochster-j1}
	\beta_{i,i+1}\!\bigl(k[\Delta]\bigr)
	=
	\sum_{\substack{W \subseteq \{x_1, \dots, x_n\} \\ |W| = i+1}}
	\left(\beta_0(\Delta_W) - 1\right),
\end{equation}
and for \(j \geq 2\), it takes the form
\begin{equation}\label{eq:Hochster-j2}
	\beta_{i,j+i}\!\bigl(k[\Delta]\bigr)
	=
	\sum_{\substack{W \subseteq \{x_1, \dots, x_n\} \\ |W| = j+i}}
	\beta_{j-1}(\Delta_W),
\end{equation}
where \(\beta_{j-1}(\Delta_W)\) denotes the \((j-1)\)-st Betti number of the homology of \(\Delta_W\). These expressions establish a direct connection between the topological invariants of the simplicial complex \(\Delta\) and the algebraic invariants of its associated Stanley--Reisner ring.

\subsubsection{f-vectors and h-vectors}
Let \(\Delta\) be a simplicial complex of dimension \(d-1\). The \emph{\(f\)-vector} of \(\Delta\) is defined as
\begin{equation}\label{eq:f-vector-def}
	(f_0, f_1, \dots, f_{d-1}),
\end{equation}
where \(f_i\) denotes the number of \(i\)-dimensional faces of \(\Delta\). By convention, we set \(f_{-1} = 1\) to account for the empty face.

The \emph{Hilbert series} of the Stanley--Reisner ring \(k[\Delta]\), also referred to as the Hilbert series of \(\Delta\), is given by
\begin{equation}\label{eq:Hilbert-series-def}
	H_{\Delta}(s)
	=
	\sum_{d \geq 0} \dim_k\bigl(k[\Delta]_d\bigr) \, s^d,
\end{equation}
where \(k[\Delta]_d\) denotes the degree-\(d\) component of the \(\mathbb{Z}\)-graded ring \(k[\Delta]\).

For a \((d-1)\)-dimensional simplicial complex \(\Delta\), it is a classical result that the Hilbert series can be expressed as a rational function of the form
\begin{equation}\label{eq:Hilbert-series-classical}
	H_{\Delta}(s)
	=
	\frac{\,h_0 + h_1 s + \cdots + h_d s^d\,}{(1 - s)^d},
\end{equation}
where \((h_0, h_1, \dots, h_d)\) is the \emph{\(h\)-vector} of \(\Delta\), or equivalently, of its Stanley--Reisner ring.

The \(f\)-vector and \(h\)-vector are related by the identity
\begin{equation}\label{eq:f-in-terms-of-h}
	\sum_{j=0}^{d} h_j\, s^j
	=
	\sum_{j=0}^{d} f_{j-1} (1 - s)^{d - j} s^j,
	\quad
	\text{with} \quad f_{-1} = 1.
\end{equation}
Equivalently, the entries of the \(h\)-vector can be expressed in terms of the \(f\)-vector by the relation
\begin{equation}\label{eq:h-in-terms-of-f}
	h_j
	=
	\sum_{i = 0}^{j}
	(-1)^{j - i} \binom{d - i}{j - i} f_{i-1},
	\quad
	j = 0, 1, \dots, d,
\end{equation}
and conversely, the \(f\)-vector can be recovered from the \(h\)-vector via
\begin{equation}\label{eq:f-h-binomial}
	f_{j-1}
	=
	\sum_{i = 0}^{j}
	\binom{d - i}{j - i} h_i,
	\quad
	j = 0, 1, \dots, d.
\end{equation}

\subsubsection{Filtration and Persistent Stanley--Reisner Theory (PSRT)}

A fundamental limitation of using a simplicial complex \(\Delta\) to model data is that it typically captures topological or combinatorial information at a single scale, thereby omitting geometric details that may vary across scales. To address this limitation, one introduces a multiscale framework through the use of filtrations, leading to the theory of persistent homology, which identifies topological features that persist across a range of scales.

Persistent Stanley--Reisner theory follows a similar philosophy, extending combinatorial and algebraic invariants to a persistent setting. Let \(\Delta\) be an abstract simplicial complex on a finite vertex set \(V\). Given a monotone function \(f \colon \Delta \to \mathbb{R}\), that is,
\[
\tau \subseteq \sigma \quad \Rightarrow \quad f(\tau) \leq f(\sigma),
\]
we define the induced filtration of \(\Delta\) by
\begin{equation}\label{eq:filtration}
	\widetilde{f} = (\Delta_f^t|t \in \mathbb{R}),
\end{equation}
where each subcomplex \(\Delta_f^t \subseteq \Delta\) is defined as
\[
\Delta_f^t := \left\{ \sigma \in \Delta \mid f(\sigma) \le t \right\}.
\]
For notational simplicity, we may write \(\Delta^t\) in place of \(\Delta_f^t\) when the context is clear.

Given a subset \(W \subseteq V\), then if \((\Delta^t)_{t \in \mathbb{R}}\) is a filtration of \(\Delta\), the induced filtration on the subcomplex \(\Delta_W\) is given by
\begin{equation}\label{eq:induced-subcomplex-filtration}
	\Delta_W^t := \Delta^t \cap \Delta_W \subseteq \Delta^{t'} \cap \Delta_W = \Delta_W^{t'}
	\quad \text{for all } t \le t'.
\end{equation}

We define the \emph{persistent Stanley--Reisner graded Betti number} \(\beta_{i, i+j}^{t,t'}(k[\Delta])\) as
\begin{align}\label{eq:persistent-Betti}
	\beta_{i, i+j}^{t,t'}\bigl(k[\Delta]\bigr) 
	&= \sum_{\substack{W \subseteq V \\ |W| = i + j}}
	\dim_k \left(
	\iota_{j-1}^{t,t'} 
	\colon 
	\widetilde{H}_{j-1}(\Delta_W^t; k) 
	\to 
	\widetilde{H}_{j-1}(\Delta_W^{t'}; k)
	\right) %\notag \\
%	&= \sum_{\substack{W \subseteq V \\ |W| = i + j}}
%	\beta_{j-1}(\Delta^t_W)
\end{align}
where \(\iota_{j-1}^{t,t'}\) is the homomorphism induced by the inclusion \(\Delta_W^t \hookrightarrow \Delta_W^{t'}\) on the \((j-1)\)-st reduced homology, and \(\beta_{j-1}(\Delta^t_W)\) is the persistent Betti number of the inclusion from \(\Delta_W^t\) to \(\Delta_W^{t'}\). Summing over all relevant subsets \(W\) yields a multiscale refinement of Hochster’s formula, capturing the persistent homological features of the complex across varying levels of filtration.

The persistent Stanley--Reisner graded Betti numbers \(\beta_{i, i+j}^{t,t'}(k[\Delta])\) generalize classical persistent Betti numbers; for example,
\[
\beta_{i, |V|}^{t,t'} = \beta_{|V| - i - 1}^{t,t'},
\]
and further encode additional combinatorial information by tracking all monomial degrees in the resolution.

One may also extend combinatorial invariants such as the \(f\)-vector and \(h\)-vector to persistent settings. The \emph{persistent \(h\)-vector} is defined by
\begin{equation}\label{persistence_h}
	h_m^{t,t'} = \sum_{j=0}^{m}
	\binom{n - d + m - j - 1}{m - j}
	\left(
	\sum_{i=0}^{j} (-1)^i \beta_{i,j}^{t,t'}
	\right),
\end{equation}
and the corresponding \emph{persistent \(f\)-vector} is given by
\begin{equation}\label{persistence_f}
	f_{m-1}^{t,t'} = \sum_{i=0}^m
	\binom{d - i}{m - i} \, h_i^{t,t'},
	\quad m = 0,1,\dots,d.
\end{equation}

Finally, consider a filtration \((\Delta^t)_{t \in \mathbb{R}}\) of the simplicial complex \(\Delta\). As \(t\) increases, the corresponding Stanley--Reisner ideals \(I(\Delta^t)\) evolve through the reverse inclusion
\[
I(\Delta^{t'}) \subseteq I(\Delta^t) \quad \text{for } t \le t'.
\]
Let \(\mathcal{P}(\Delta^t)\) denote the set of facet prime monomial ideals (or \emph{facet ideals}) of \(\Delta^t\). For each \(i \ge 0\), define \(\mathcal{P}_i(\Delta^t)\) to be the subcollection of \(\mathcal{P}(\Delta^t)\) consisting of those associated to \(i\)-dimensional facets. Then, we have the disjoint union
\[
\mathcal{P}(\Delta^t) = \bigsqcup_{i=0}^{\dim(\Delta^t)} \mathcal{P}_i(\Delta^t).
\]
This decomposition motivates a persistent interpretation of the evolution of facet ideals over the filtration. In analogy with persistent homology, we refer to the facet ideals \(P_\sigma\) of \(I(\Delta^t)\) as the \emph{persistent facet ideals} of \(\Delta\), representing combinatorial structures that persist across filtration levels.

We define the \emph{facet persistence betti number} \(\beta_i^{t,t'}\) to be the number of the persistent facet ideals in \(\mathcal{P}_i(\Delta^t)\) that are persistent facet ideals in \(\mathcal{P}_i(\Delta^{t'})\). Explicitly,
\[
\beta_i^{t,t'} = |\mathcal{P}_i(\Delta^t)\cap \mathcal{P}_i(\Delta^{t'})|.
\]

\begin{figure}[h!]
		\centering
		\subfigure{
			\includegraphics[width=0.88\textwidth]{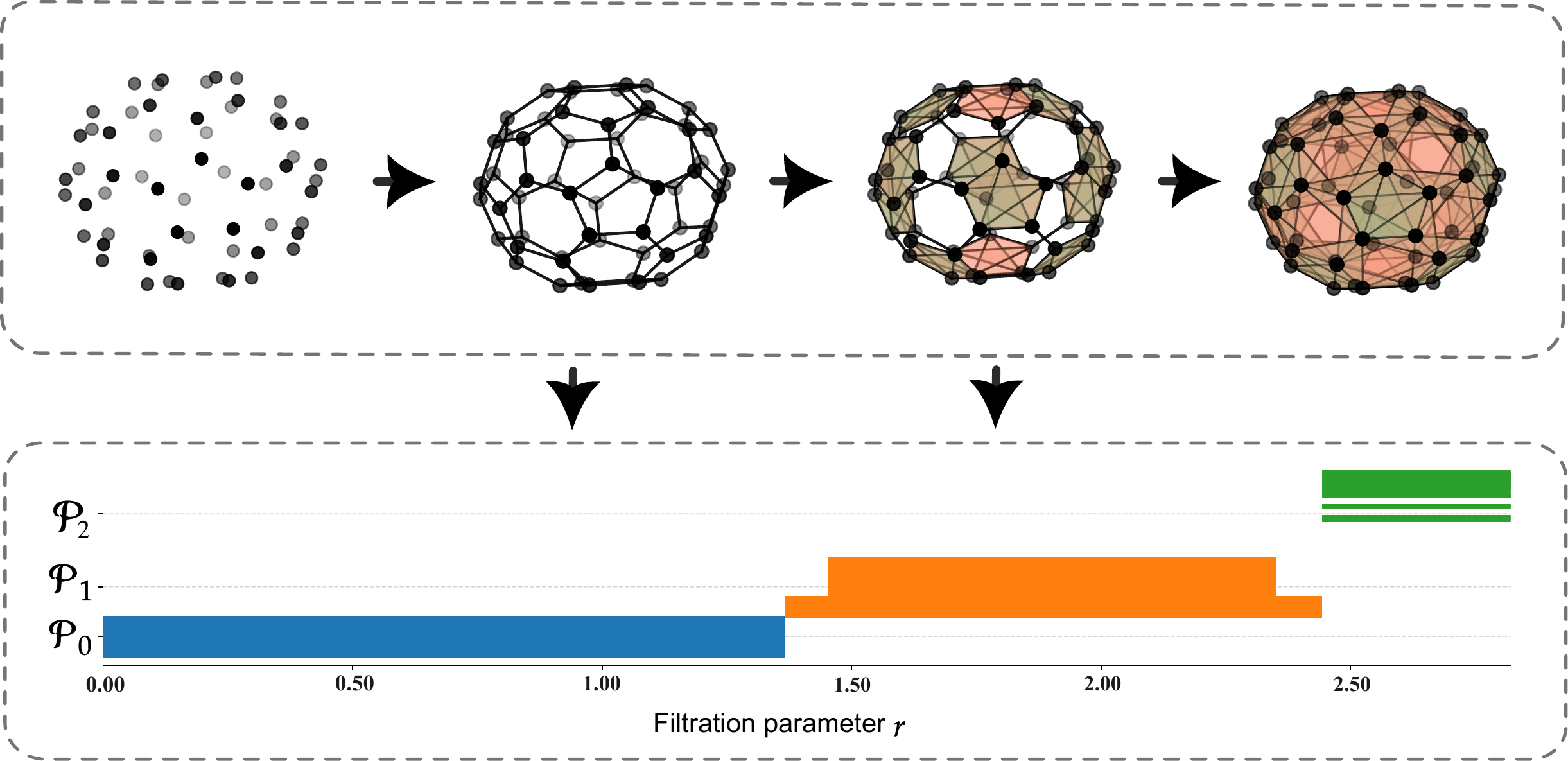} % Path to Figure 1
		}
		\caption{Illustration of the filtration process of the persistent commutative algebra. Given a point cloud input (a C$_{60}$ molecule), a corresponding simplicial complex with an associated filtration is generated. Critical values and facet persistence barcodes are computed in various dimensions.
		}
		\label{fig:scheme}
	\end{figure}

\subsection{Vectorization of persistent comutative algebra}\label{sec:Vectorization}

\paragraph{Commutative algebra on bipartite complexes}

In (metallo)protein-ligand binding interactions, bipartite complexes refer to molecular assemblies composed of two distinct, non-overlapping (metallo)protein and ligand components that interact to form a functional unit. In graph theory, a bipartite complex refers to a structure where vertices (or nodes) are divided into two distinct, disjoint sets, and edges only connect vertices from different sets. This bipartite structure is adopted in our commutative algebra analysis of (metallo)protein-ligand binding data.

\paragraph{Element-specific commutative algebra.}

There are various types of intramolecular and intermolecular interactions, including hydrogen bonding, electrostatic forces, and hydrophobic and hydrophilic interactions. To effectively characterize these critical interactions, element-specific modeling was developed in our earlier work \cite{cang2017topologynet}, demonstrating notable effectiveness. In particular, molecular interactions enriched within various pairwise combinations of atom sets are captured using persistent commutative algebra (PCA) modeling and subsequently represented through PCA-based vectorization.

For the PDBbind-v2016 dataset, molecular interactions are characterized based on four commonly occurring atom types in proteins: carbon (C), nitrogen (N), oxygen (O), and sulfur (S), along with ten atom types in ligands, including carbon (C), nitrogen (N), oxygen (O), sulfur (S), phosphorus (P), fluorine (F), chlorine (Cl), bromine (Br), iodine (I), and hydrogen (H). In the case of the metalloprotein–ligand dataset, additional atomic interactions involving metal ions must be considered. Based on statistical analysis of the metal ions frequently occurring in protein–ligand binding pockets, we include seven additional atom types: zinc (Zn), magnesium (Mg), manganese (Mn), calcium (Ca), sodium (Na), iron (Fe), and nickel (Ni), due to their prevalent presence.

For protein--ligand complexes in the PDBbind-v2016 dataset, it is sufficient to consider only protein--ligand interactions. However, for metalloprotein--ligand binding affinity prediction, three types of interactions must be taken into account: protein--ligand interactions, metal ion--ligand interactions, and protein--metal ion interactions. Specifically, there are $4 \times 10 = 40$ atom pair combinations for protein--ligand interactions (P--L), $7 \times 4 = 28$ combinations for metal ion--protein interactions (M--P), and $7 \times 10 = 70$ combinations for metal ion--ligand interactions (M--L). Consequently, $40$ types of atomic interactions are considered for protein--ligand complexes in the PDBbind-v2016 dataset, while a total of $138$ interaction types are incorporated for metalloprotein--ligand complexes.

For the 3D coordinate point cloud corresponding to each atom group combination, we utilize PCA to generate vectorized representations. The persistent facet Betti numbers and their corresponding rates are employed to design a set of features for each atom group. A cutoff distance of 12~\AA~from the ligand is used to collect nearby protein atoms, while a cutoff distance of 15~\AA~from the ligand is applied to gather metal atoms. In the PCA-based featurization process, the filtration range is set from 1 to 12~\AA, with a step size of 0.5~\AA~for the filtration steps.

Rips complexes are constructed for each point cloud, while bipartite complexes are generated to capture the three types of molecular interactions. In our implementation, we focus exclusively on the facet Betti numbers and their corresponding rates for 0-simplices and 1-simplices. The final molecular descriptor is obtained by concatenating the PCA-derived feature vectors from each point cloud. \autoref{fig:scheme} presents a schematic illustration of the general persistent facet Betti curves computed for a representative set of atoms.

\paragraph{Category-specific commutative algebra.}

In addition to element-specific modeling, we also adopt a category-specific strategy to capture intrinsic molecular interactions characterized by amino acid types. Proteins are composed of 20 common amino acid residues, which can be classified into four major categories based on their side chain properties: hydrophobic (H), uncharged (U), negatively charged (N), and positively charged (P). We denote the atoms belonging to these categories as $A_H$, $A_U$, $A_N$, and $A_P$, respectively. \autoref{tab:categories} summarizes the four categories along with their corresponding amino acid types. 

As with the element-specific atom groupings used for ligands and metal ions, we consider 10 atom groups in ligands and 7 atom groups in metal ions based on their element types. Consequently, there are 40 atom group combinations in general protein–ligand complexes, and 138 combinations in metalloprotein–ligand complexes. Similar PCA-based vectorizations are applied to these atom group combinations to generate molecular descriptors.

\begin{table}[htb!]
	\small
	\centering
	\begin{tabular}{c |c |c |c }
		\hline
		\textbf{Hydrophobic (H)} &  \textbf{Uncharged (U)} &  \textbf{Negatively Charged (N)} & \textbf{Positively Charged (P)}\\
		\hline
		Glycine      (Gly) & Serine	    (Ser)& Aspartic (Asp)&Lysine	(Lys) \\
		Alanine      (Ala) & Threonine  (Thr)& Glutamic (Glu)&Arginine (Arg)\\
		Valine	     (Val) & Asparagine (Asn)&               &Histidine     (His)\\
		Leucine      (Leu) & Glutamine  (Gln)&               & \\
		Isoleucine   (Ile) & Tyrosine   (Tyr)&               & \\
		Methionine   (Met) & Cysteine   (Cys)&               & \\
		Proline      (Pro) &            &               & \\
		Phenylalanine(Phe) &            &               & \\
		Tryptophan   (Trp) &            &               & \\
		\hline
	\end{tabular}
	\caption{The four categories of amino acid residues in proteins according to their side chain properties.}
	\label{tab:categories}
\end{table}

 \subsection{Natural language processing (NLP) molecular descriptors}
 
Natural language processing (NLP) recently also become popular machine-learning techniques for molecular biosciences. We utilize NLP techniques to boost the performance of our PSRT-based machine learning models for protein-ligand binding affinity predictions. Different from our PSRT theory that analyze molecular 3D structures, NLP  extract molecular physichemical properties by analyzing molecular sequence patterns. For protein-ligand complex, we have amino acid sequences for the protein and SMILES strings for the ligand. Some NLP-based molecular descriptors are designed using transformer techniques for both protein and ligand in works \cite{rives2021biological,chen2021extracting}. By concatenation molecular descriptors from those pretrained deep learning models for protein sequence and ligand SMILES strings, we obtained a sequence representation for protein-ligand complex.
 
 \subsubsection{ESM transformer protein language model}
 
The ESM-2 transformer model, introduced by Rives et al.~\cite{rives2021biological}, has become one of the most widely adopted protein language models, with applications in protein engineering and drug discovery. This model was trained on a dataset containing 250 million amino acid sequences and employs a deep learning architecture with 34 layers and 650 million parameters. In this work, we utilized the ESM-2 model to generate sequence embeddings for proteins. At each layer, a sequence of length~\(L\) is encoded into a matrix of size~\(1280 \times L\), excluding the start and end tokens. We extracted the sequence representation from the final (34th) layer and computed the average along the sequence length axis, resulting in a 1280-dimensional feature vector.

 \subsubsection{Transformer-based small molecular language model}
 
 A transformer-based deep learning framework was introduced to extract molecular representations~\cite{chen2021extracting}, serving as a powerful tool for machine learning applications involving small molecules~\cite{shen2023svsbi}. This model was trained on a collection of over 700 million SMILES strings obtained from databases such as ChEMBL, PubChem, and ZINC. Three pretrained variants were developed: model-C, model-CP, and model-CPZ. In the current study, we utilize the model-CPZ to generated molecular descriptors for ligands. For each ligand, the model produces a matrix of size~\(256 \times 512\), where 256 corresponds to the symbols representing the molecule and 512 is the dimension of the embedding vector for each symbol. The final molecular descriptors are obtained by first vector among the 256 embedding vectors, resulting in a fixed-length feature vector.

\subsection{Machine learning modeling}

We employ the Gradient Boosting Decision Tree (GBDT) algorithm to develop our machine learning models, using the Python \texttt{scikit-learn} package (v1.3.2) for implementation. GBDT is well-regarded for its robustness against overfitting, relative insensitivity to hyperparameter settings, and ease of implementation.  The algorithm creates multiple weak learners or individual trees by bootstrapping training samples and integrates their outputs to make predictions. Although weak learners are prone to making poor predictions, the ensemble approach can reduce overall errors by combining the predictions of all the weaker learners. We input resulting PSRT molecular descriptors and transformer-based molecular dscriptors into GBDT algorithm to build regression models, respectively. The GBDT hyperparameters used for modeling are listed in \autoref{table:GBDT-parameters}.

\begin{table}[htb!]
	\small
	\centering
	\begin{tabular}{c c c  c }
		\hline
		No. of estimators &  Max depth & Min. sample split & Learning rate\\
		20000/30000& 7 &5 & 0.002\\
		\hline
		Max features & Subsample size & Repetition &\\
		Square root & 0.8&  20 times & \\
		\hline
	\end{tabular}
	\caption{Hyperparameters used for build gradient boosting regression models. Tree numbers are set to be 20000 and 30000 respectively for PSRT and transformer-based molecular descriptor modeling.}
	\label{table:GBDT-parameters}
\end{table}

\subsection{Evaluation metrics}

To quantitatively evaluate the performance of our binding affinity prediction models, we employ the Pearson correlation coefficient (PCC), defined as:
\begin{align*}
	\text{PCC}(\mathbf{x}, \mathbf{y}) = \frac{\sum_{m=1}^{M} (y_m^e - \bar{y}^e)(y_m^p - \bar{y}^p)}{\sqrt{\sum_{m=1}^{M} (y_m^e - \bar{y}^e)^2 \sum_{m=1}^{M} (y_m^p - \bar{y}^p)^2}},
\end{align*}
where \( y_m^e \) and \( y_m^p \) denote the experimental and predicted binding affinity values for the \( m \)-th sample, respectively, and \( \bar{y}^e \) and \( \bar{y}^p \) are their corresponding mean values.

We also report the root mean squared error (RMSE), which is computed as:
\begin{align*}
	\mathrm{RMSE} = \sqrt{\frac{1}{n} \sum_{m=1}^{M} (y_m^e - y_m^p)^2},
\end{align*}
where \( y_m^e \) and \( y_m^p \) represent the experimental and predicted binding affinity values for the \( m \)-th sample, respectively.

We employ the above two metrics to assess the performance of our machine learning models on both datasets. The original labels for these datasets are given as \( pK_d \) values, which can be converted to binding free energies (in kcal/mol) by multiplying by a constant factor of 1.3633. Our models achieve low RMSE values across both datasets. For the PDBbind-v2016 dataset, we convert the labels to binding energies and use them for RMSE comparisons with previously published models.

\section{Conclusion}
As a foundational part of algebraic geometry and algebraic number theory, commutative algebra studies commutative rings, their ideals, and modules over such rings. However, commutative algebra has rarely been applied to data science and machine learning.
The persistent Stanley-Reisner theory (PSRT) \cite{suwayyid2025persistent}, introduced by Suwayyid and Wei, offers a new opportunity to develop commutative algebra machine learning (CAML) and commutative algebra deep learning (CADL) for data. Stanley–Reisner theory, also known as face ring theory, creates a profound connection between combinatorics and commutative algebra. PSRT integrates tools from algebra, combinatorics, and multiscale analysis (i.e., filtration) to study simplicial complexes via Stanley–Reisner rings.

This work proposes CAML for data analysis. We pair PSRT with a robust machine learning method, gradient boosted decision trees (GBDT), which utilizes an ensemble of decision trees to make predictions. GBDT is known for its high accuracy and efficiency, particularly for relatively small datasets that are not suitable for deep learning algorithms. We consider two biomolecular datasets, i.e., a protein-ligand binding dataset (PDBbind-v2016) and a metalloprotein-ligand binding dataset, to validate the proposed CAML model. Due to the intricate interactions in (metallo)protein-ligand complexes, we propose new algorithms, such as commutative algebra on bipartite complex, element-specific commutative algebra, and category-specific commutative algebra, to capture the physics and chemistry underlying the interactions. The performance of the proposed CAML model is compared with other state-of-the-art methods in the literature. We demonstrate that CAML is an extremely promising new method for protein-ligand binding predictions.

Protein-ligand binding prediction serves as a case study of the proposed CAML model. CAML can easily be applied to other biomolecular data predictions and problems in science and engineering. We believe that CAML represents an emerging direction in machine learning and data science.

 \section*{Data Availability}
  
  All data and the code needed to reproduce this paper's result can be found at\\ \href{https://github.com/WeilabMSU/CAML}{https://github.com/WeilabMSU/CAML}. For detailed information on the metalloprotein–ligand complex dataset, please refer to the reference \cite{jiang2023metalprognet}. The PDBbind-v2020 dataset is available at \url{http://pdbbind.org.cn/}.

\section*{Acknowledgments}
This work was supported in part by NIH grants   R01AI164266, and R35GM148196, NSF grant  DMS-2052983,    MSU Foundation, and Bristol-Myers Squibb 65109.
F.S. thanks King Fahd University of Petroleum and Minerals for their support.
  C.-L. C. gratefully acknowledges financial support from the Defense Threat Reduction Agency (Project CB11141), and the Department of Energy (DOE), Office of Science, Office of Basic Energy Sciences (BES) under an award FWP 80124 at Pacific Northwest National Laboratory (PNNL). PNNL is a multiprogram national laboratory operated for the Department of Energy by Battelle under Contract DE-AC05-76RL01830.

%\clearpage

%\bibliographystyle{unsrt}
%\bibliography{refs}

\end{document}